# 2D Addressable Mid-infrared Metasurface Spatial Light Modulator

Cosmin-Constantin Popescu[1], Maarten Robbert Anton Peters[1], Oleg Maksimov[2], Harish Bhandari[2], Rashi Sharma[3], Kathleen Richardson[3], Arka Majumdar[4], Hyun Jung Kim[5], Rui Chen[1], Khoi Phuong Dao[1], Luigi Ranno[1], Brian Mills[1,6], Dennis Calahan[7], Tian Gu[1,8], and Juejun Hu[1,8,*]

[1]Department of Materials Science & Engineering, Massachusetts Institute of Technology, Cambridge, MA, USA
[2]Radiation Monitoring Devices, Inc., Watertown, MA, USA
[3]The College of Optics & Photonics, Department of Materials Science and Engineering, University of Central Florida, Orlando, FL, USA
[4]Department of Electrical and Computer Engineering, University of Washington, Seattle, WA, USA
[5]The Department of Aerospace Engineering, KAIST, Daejeon, South Korea
[6]Draper Scholar Program, The Charles Stark Draper Laboratory, Inc., Cambridge, MA, USA
[7]The Charles Stark Draper Laboratory, Inc., Cambridge, MA, USA
[8]Materials Research Laboratory, Massachusetts Institute of Technology, Cambridge, MA, USA

*hujuejun@mit.edu

**Abstract**

Active metasurfaces enable dynamic control of light for applications in beam steering, pixelated holography, and adaptive optics, but demonstrations of two-dimensional (2D) electrically addressable arrays have so far been limited. Here we introduce a scalable 2D architecture based on phase-change materials (PCMs) integrated metasurfaces and apply it to realize the first transmissive mid-infrared (mid-IR) spatial light modulator (SLM). The device is fabricated through standard silicon photonic foundry processing combined with backend-of-line (BEOL) integration and employs multilayer backend metal interconnects to implement a crossbar addressing scheme. Each pixel is integrated with a silicon diode selector to suppress sneak-path currents, a feature essential for scaling to large arrays. The result establishes a foundry-compatible route to high-density, large-area active metasurfaces with independently tunable pixels.

## Introduction

Active metasurfaces have emerged as a powerful platform for dynamically controlling light, enabling functions such as beam steering, holography, adaptive focusing, and on-demand wavefront shaping[1–7]. The most versatile implementations require pixel-addressable tuning, in contrast to global tuning approaches that modulate all meta-atoms simultaneously[8]. Pixel-level control is essential for two reasons. First, it allows continuous tuning of the optical response, for example beam steering over a continuous range of angles or varifocal lenses with smoothly adjustable focal lengths, rather than being restricted to discrete switching between globally defined states. Second, it enables spatially dependent modulation of amplitude, phase, and polarization across the device aperture, which, in the limit of single-meta-atom control, can realize so-called universal optics: a reconfigurable platform capable of implementing arbitrary linear optical functions[9]. While one-dimensional electrical addressing can be realized with straightforward fan-out architectures[10–17], extending this approach to two dimensions is only feasible for very small arrays containing a handful of independently controlled pixels (Table 1)[18–20].

Here we present a scalable 2D electrically addressable active metasurface architecture that overcomes the scaling bottleneck. The design combines BEOL-integrated PCMs with foundry-processed silicon microheater arrays addressed through multilayer backend metal interconnects, eliminating the need for complex fan-out wiring and avoiding interconnect congestion as array size increases. Importantly, each pixel incorporates a silicon diode selector to eliminate sneak-path currents, enabling independent pixel control across large arrays.

We implement this active metasurface platform to address a critical gap in SLM technology: the absence of transmissive devices operating in the mid-infrared. Existing SLM platforms, such as liquid-crystal-on-silicon (LCoS) devices, suffer from strong absorption of liquid crystals in the mid-IR and relatively slow response times[21]. Digital micromirror devices (DMDs) operate only in a reflective geometry and primarily provide amplitude modulation[22]. As a result, transmissive mid-IR SLMs remain effectively unexplored, despite their advantages for compact optical layouts, cascaded optical processing, and integration with transmissive sensing or imaging systems. Our approach directly fills this gap. By leveraging the broadband transparency of PCMs[23], the same architecture can be adapted to operate across spectral bands from the near- to the long-wavelength infrared. The PCM metasurface platform is inherently versatile, capable in principle of modulating phase, amplitude, polarization, or even optical angular momentum[24–31]. The use of nonvolatile PCMs enables persistent states without static power consumption, and reliance on foundry manufacturing provides a clear path toward continued performance scaling in pixel count, aperture size, and modulation complexity.

## 2D electrically addressable active metasurface matrix

The metasurface pixel architecture is illustrated in Fig. 1a. The crossbar fabric containing the pixel array was fabricated using AIM Photonics' standard Active Silicon process, which provides two BEOL metal layers used to form the orthogonal row and column addressing lines. At each crosspoint, the intersecting wires connect to a doped-silicon microheater in series with a Si PIN diode. Both the diodes and heaters are implemented using the multiple doping levels available in the commercial AIM Photonics process. The diodes serve as unipolar selectors, blocking sneak-path currents as depicted in Fig. 1a—a critical function for large crossbar arrays, where sneak-path current otherwise increases rapidly with array size. Quantitative analysis in Supplementary Section I underscores the impact of selector integration: in an array without selectors, similar to prior demonstrations[32], sneak-path currents already consume more power than the addressed pixel in a modest 4 × 4 array. By contrast, our selector-enabled architecture can scale to megapixel arrays with negligible parasitic power loss. The ability to seamlessly integrate pixel-level selectors is a unique feature enabled by Si foundry fabrication, alongside the inherent manufacturing scalability of the process. The nonvolatile nature of PCM further simplifies the architecture: instead of requiring a per-pixel transistor, a Si diode integrated with each microheater enables single-pixel addressing and sequential row- or column-based programming.

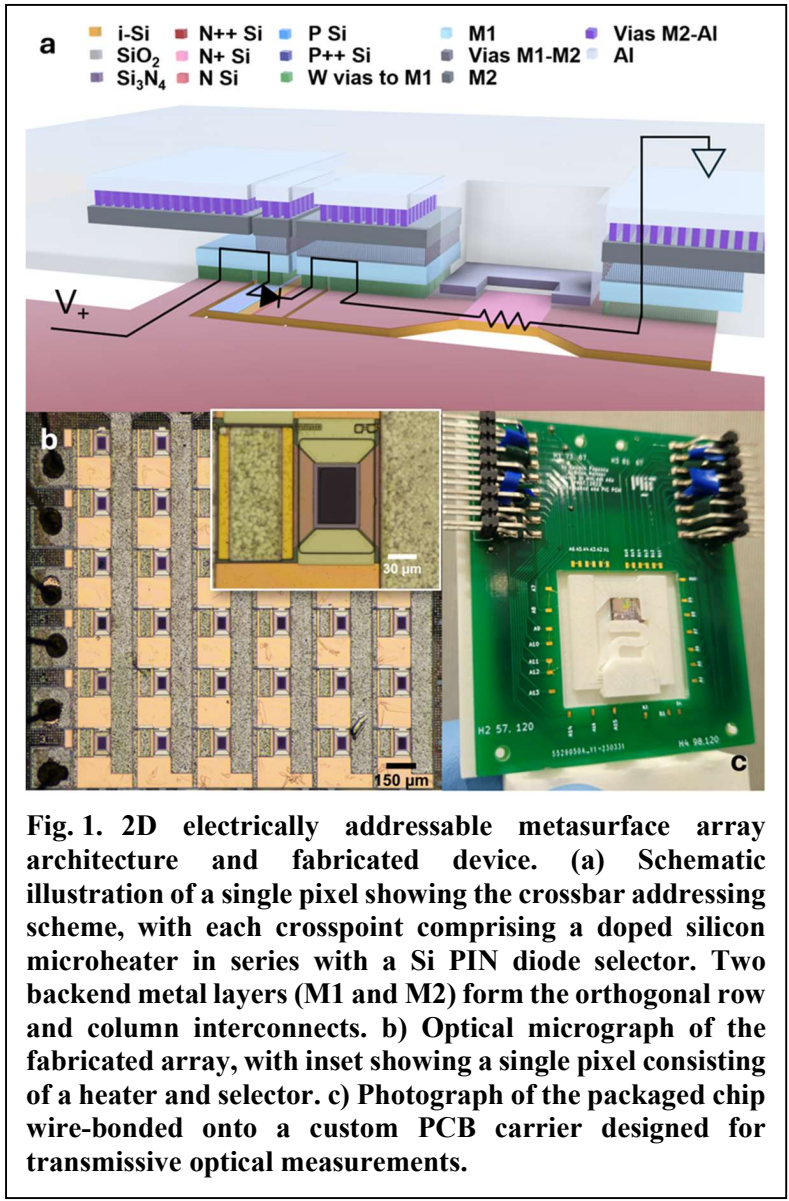

**Fig. 1. 2D electrically addressable metasurface array architecture and fabricated device. (a)** Schematic illustration of a single pixel showing the crossbar addressing scheme, with each crosspoint comprising a doped silicon microheater in series with a Si PIN diode selector. Two backend metal layers (M1 and M2) form the orthogonal row and column interconnects. **b)** Optical micrograph of the fabricated array, with inset showing a single pixel consisting of a heater and selector. **c)** Photograph of the packaged chip wire-bonded onto a custom PCB carrier designed for transmissive optical measurements.

Following foundry fabrication, windows are opened in the BEOL dielectric to expose the microheaters. In this step, the SiN layer available as part of the AIM Photonics process serves as an etch stop, ensuring minimal damage to the underlying Si heaters. Electrical measurements performed before and after window opening confirm that heater performance is preserved (Supplementary Section II). $Ge_2Sb_2Se_4Te$ (GSST) films are deposited into these windows and then patterned into metasurface pixels via electron beam lithography and plasma etching. The pixels are encapsulated with a two-layer capping structure, which as shown in our previous study[33] significantly improves device endurance. Optical micrographs of the fabricated array and a single pixel are shown in Fig. 1b. The completed chips are wire-bonded to a custom printed circuit board (PCB) carrier designed for transmissive optical measurements (Fig. 1c). Fabrication and packaging details are given in Methods and Supplementary Section II.

**Metasurface pixel design and characterization**

The metasurface pixel design, shown in Fig. 2a, is based on the principle of guided-mode resonance (GMR). A phase transition in the PCM changes the effective index of the guided mode, producing a shift in the resonance wavelength and modulating the transmission amplitude. In addition to supporting GMR, the "fishnet" metasurface design provides an array of anchoring points where the bilayer capping structure makes direct contact with the Si heater. This mechanical linkage helps mitigate PCM delamination as a leading cause of PCM device failure and thereby contributes to improved device endurance[34]. Fig. 2b presents rigorous coupled-wave analysis

(RCWA) simulations of the pixel transmission spectra in the amorphous (Am) and crystalline (Cr) states for light linearly polarized along the *y*-direction, showing that refractive-index contrast between the two phases modulates the GMR wavelength and generates strong optical contrast. By tuning the coupling strength to operate near the critical coupling regime, the design predicts a switching contrast ratio of 15 dB near the crystalline-state GMR wavelength of 2.65 μm. The GMR wavelength exhibits distinctive dependences on the metasurface lattice periods along the *x* and *y* axes, $p_x$ and $p_y$: varying $p_x$ provides an effective means of tuning the resonance, whereas the wavelength is largely insensitive to changes in $p_y$, as illustrated in Fig. 2c and Fig. 2d. Details of the simulation protocols and results are elaborated in Supplementary Section III.

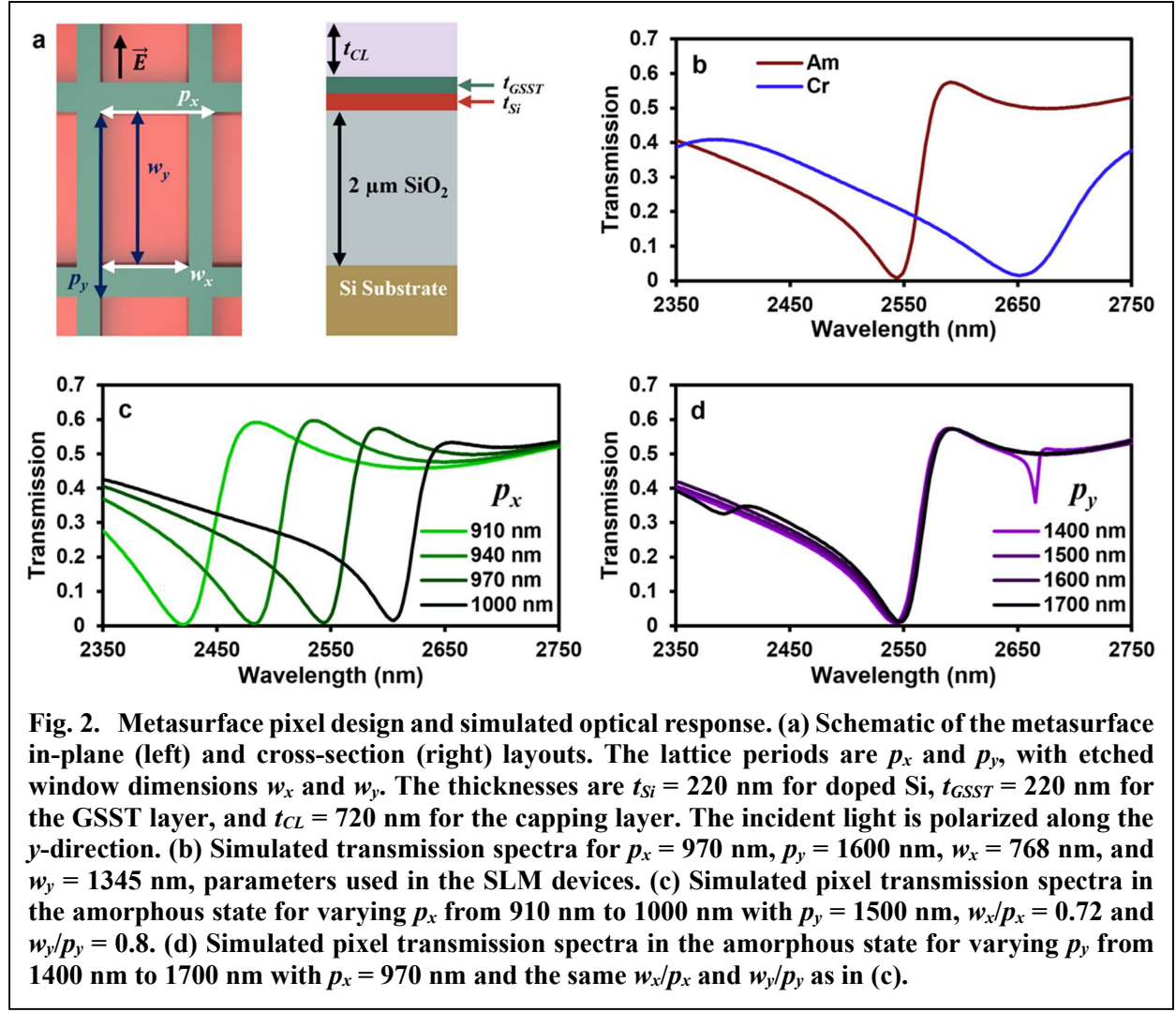

**Fig. 2. Metasurface pixel design and simulated optical response.** (a) Schematic of the metasurface in-plane (left) and cross-section (right) layouts. The lattice periods are $p_x$ and $p_y$, with etched window dimensions $w_x$ and $w_y$. The thicknesses are $t_{Si}$ = 220 nm for doped Si, $t_{GSST}$ = 220 nm for the GSST layer, and $t_{CL}$ = 720 nm for the capping layer. The incident light is polarized along the *y*-direction. (b) Simulated transmission spectra for $p_x$ = 970 nm, $p_y$ = 1600 nm, $w_x$ = 768 nm, and $w_y$ = 1345 nm, parameters used in the SLM devices. (c) Simulated pixel transmission spectra in the amorphous state for varying $p_x$ from 910 nm to 1000 nm with $p_y$ = 1500 nm, $w_x/p_x$ = 0.72 and $w_y/p_y$ = 0.8. (d) Simulated pixel transmission spectra in the amorphous state for varying $p_y$ from 1400 nm to 1700 nm with $p_x$ = 970 nm and the same $w_x/p_x$ and $w_y/p_y$ as in (c).

Fig. 3a shows a scanning electron microscope (SEM) image of a fabricated PCM metasurface prior to capping-layer encapsulation. Fig. 3b presents the measured transmission spectra of a metasurface pixel designed with the same parameters as in Fig. 2b, in both amorphous and crystalline states. The experimental spectra exhibit a systematic blue shift of the GMR wavelengths relative to design values and reduced extinction ratios, which we attribute to slight deviations between the actual and assumed refractive indices of the PCM and doped-Si layers. The measured transmission contrast ratio reaches 5.1 dB at 2.585 μm wavelength and can be further enhanced by optimizing the design toward the critical coupling condition. We also characterized the transmission response of pixels fabricated with varying periods $p_x$ and $p_y$, with results shown in Fig. 3c and Fig. 3d. The resonance wavelength exhibits a much stronger dependence on $p_x$ than on $p_y$, in excellent agreement with the simulation trends in Fig. 2c and Fig. 2d.

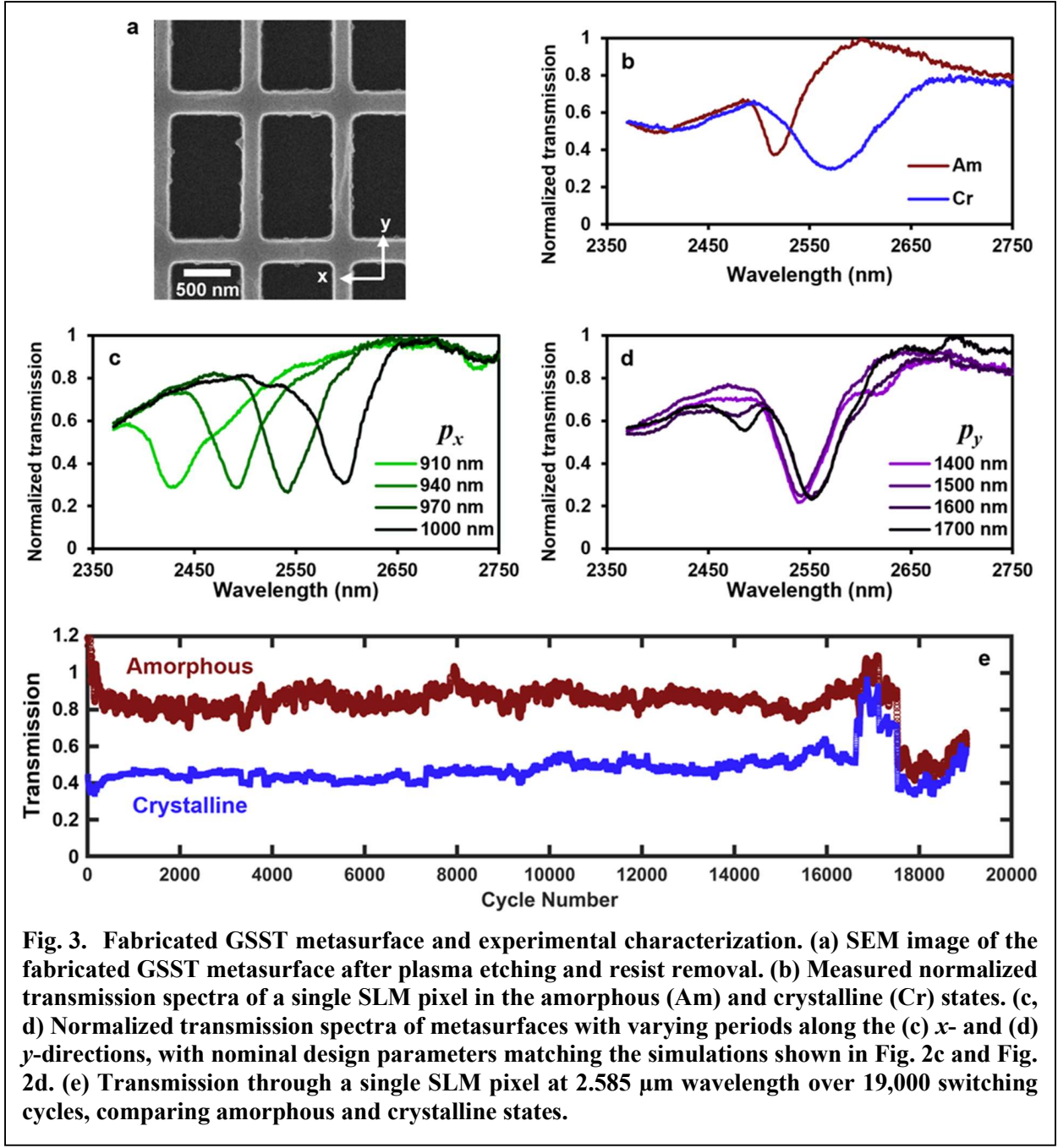

**Fig. 3. Fabricated GSST metasurface and experimental characterization. (a) SEM image of the fabricated GSST metasurface after plasma etching and resist removal. (b) Measured normalized transmission spectra of a single SLM pixel in the amorphous (Am) and crystalline (Cr) states. (c, d) Normalized transmission spectra of metasurfaces with varying periods along the (c) x- and (d) y-directions, with nominal design parameters matching the simulations shown in Fig. 2c and Fig. 2d. (e) Transmission through a single SLM pixel at 2.585 μm wavelength over 19,000 switching cycles, comparing amorphous and crystalline states.**

To toggle between the amorphous and crystalline states, we applied voltage pulses of 25.5 V for 13 μs for amorphization and 18 V for 11 ms for crystallization. These parameters were first optimized on a single pixel and subsequently applied uniformly across the entire array. The fact that the same set of pulse conditions produced reliable switching across all pixels without further adjustment highlights the excellent pixel-to-pixel uniformity and fabrication repeatability of the platform. Endurance testing results are shown in Fig. 3e, where the optical contrast was monitored over 19,000 switching cycles. The device maintained a stable switching response for approximately 16,700 cycles, after which delamination failure occurred. This endurance substantially surpasses previous reports for electrically switched PCM metasurface devices, including transmissive metasurface filters limited to ~ 1,250 cycles before degradation[33] and other metasurface implementations typically restricted to a few thousand cycles or less[35–38]. The demonstrated improvement marks a significant advance in PCM metasurface durability, with further gains anticipated through the incorporation of wetting layers to enhance adhesion and suppress delamination.

**Metasurface 2D SLM demonstration**

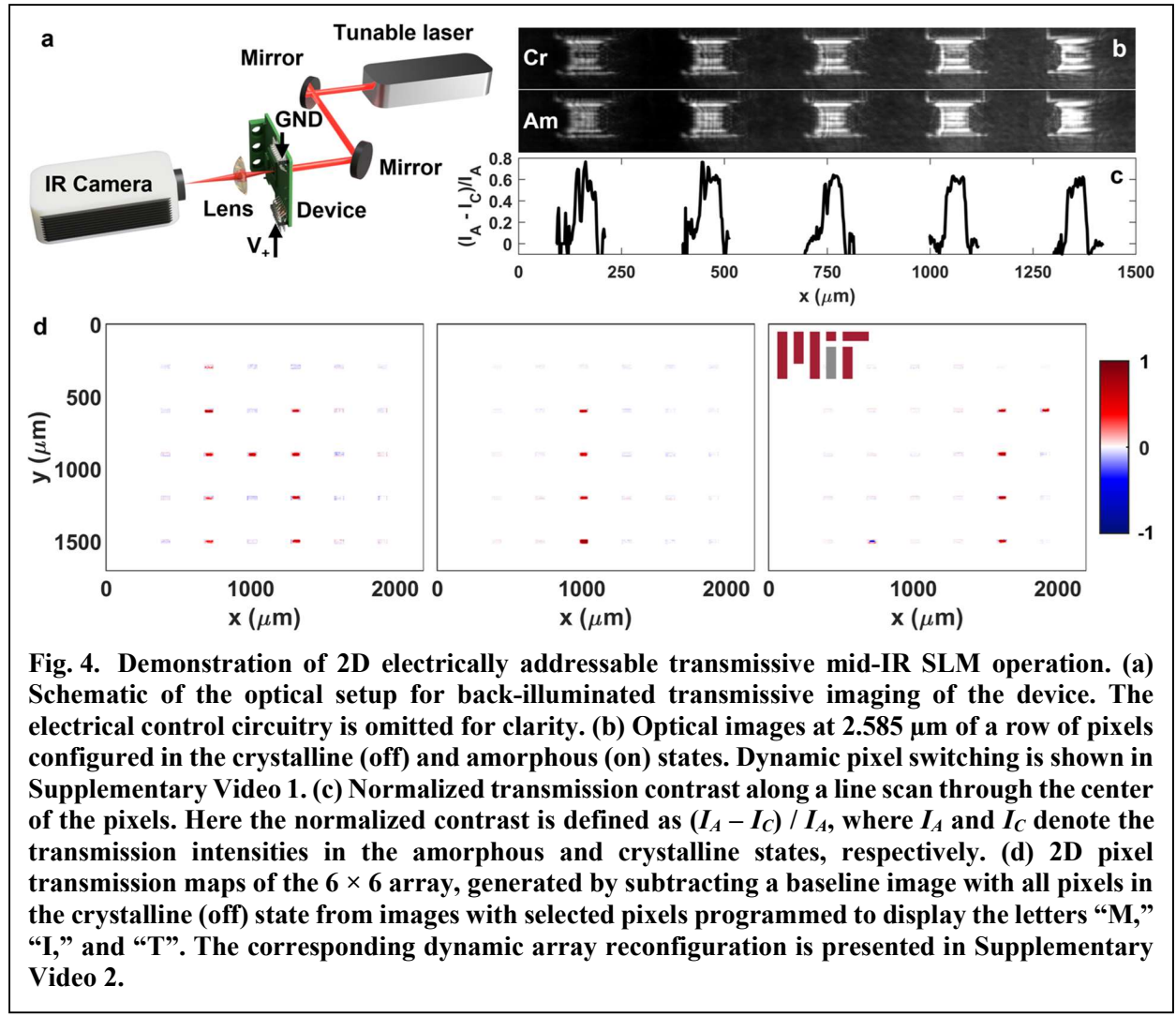

Fig. 4. Demonstration of 2D electrically addressable transmissive mid-IR SLM operation. (a) Schematic of the optical setup for back-illuminated transmissive imaging of the device. The electrical control circuitry is omitted for clarity. (b) Optical images at 2.585 μm of a row of pixels configured in the crystalline (off) and amorphous (on) states. Dynamic pixel switching is shown in Supplementary Video 1. (c) Normalized transmission contrast along a line scan through the center of the pixels. Here the normalized contrast is defined as $(I_A - I_C) / I_A$, where $I_A$ and $I_C$ denote the transmission intensities in the amorphous and crystalline states, respectively. (d) 2D pixel transmission maps of the 6 × 6 array, generated by subtracting a baseline image with all pixels in the crystalline (off) state from images with selected pixels programmed to display the letters "M," "I," and "T". The corresponding dynamic array reconfiguration is presented in Supplementary Video 2.

We implemented a 6 × 6 pixel array following the architecture in Fig. 1a, with each pixel independently addressable via electrical control. The array pitch is 400 μm, and the effective switching area per pixel is 60 × 20 μm. While the fill factor in this prototype is relatively low, it is not a fundamental limitation of the architecture. With straightforward design adjustments, the fill factor can be increased to 80-90%, values comparable to those achieved in state-of-the-art SLMs (Supplementary Section IV). Likewise, both array size and pitch can be readily optimized by leveraging the inherent scalability of the foundry fabrication process.

The SLM array was characterized using the setup shown in Fig. 4a, with further details provided in Methods. Measurements were performed at a wavelength of 2.585 μm. Fig. 4b compares two images of the array in which an entire row of metasurface pixels was set to the "on" and "off" states, respectively. The corresponding line-scan profile of the transmission contrast in Fig. 4c shows minimal pixel-to-pixel variation in either state, underscoring the uniformity of the device. Light transmission through doped-silicon regions without PCM films produces a bright static background surrounding the active switching areas of each pixel; this artifact can be readily eliminated in future implementations by adding a metal mask to block the background. Fig. 4d presents transmission maps obtained by subtracting a baseline image with all pixels in the crystalline ("off") state from images in which selected pixels were programmed to display the letters "M," "I," and "T", emulating the MIT logo shown in the inset. Supplementary Videos demonstrate dynamic switching of these programmable patterns. These results highlight the capability for on-demand, spatially reconfigurable control of the array.

**Discussion and Conclusion**

This work establishes multiple first-of-their-kind advances in active metasurface technology. To our knowledge, it represents the first report of a transmissive mid-IR SLM, the first realization of a 2D electrically addressable active metasurface with a scalable architecture, and the first report of a PCM-based metasurface fabricated using a standard silicon foundry process with full BEOL integration. These firsts are significant not only as technical milestones, but because the foundry platform inherently enables features critical to scalability: seamless integration of diode selectors at the single-pixel level, multilayer BEOL metals for crossbar routing without congestion, and multiple doping levels for designing large-area, efficient heaters, all with the uniformity and reproducibility required for consistent pixel performance. Although the present prototype is a small array with relatively large pitch and modest fill factor, the underlying architecture is directly extensible to high-density arrays. Additionally, the demonstration of extended endurance enabled by the choice and optimization of capping materials and other advances, are readily scalable, suggesting promise in the fabrication of larger arrays for a range of optical functionality. We therefore anticipate that, with straightforward design refinements, the technology will rapidly evolve to deliver performance comparable to state-of-the-art near-infrared SLMs, as supported by our projections in Supplementary Section IV.

While the present demonstration establishes multiple performance benchmarks for PCM-based metasurfaces, several aspects of the prototype leave room for improvement. First, although our device already exceeds previously reported endurance values for electrically switched PCM metasurfaces, it still falls short of the lifetimes achieved by existing SLM technologies and the requirements of many practical applications. This limitation is not intrinsic to PCMs: for example, optical switching in $Ge_2Sb_2Te_5$ has achieved endurance of $10^8$ cycles[39]. The comparatively lower endurance of electrically switched devices is generally attributed to greater non-uniformity and localized heating, but our recent work (to be published) has demonstrated over $10^7$ cycles in PCM-integrated waveguides using electrically driven heaters. We anticipate that incorporating designs such as inverse-designed microheaters with improved thermal uniformity[40] can deliver similar or superior endurance in metasurface platforms.

Second, while GSST offers broadband transparency in both phases extending beyond 20 μm[23], practical operation is constrained by absorption in the buried oxide and doped-silicon heater. The former introduces significant loss beyond ~ 7.5 μm, which could be mitigated by backside oxide removal (at the expense of mechanical stability) or replacement with lower-loss dielectrics such as SiN. Losses from doped Si can be reduced by lowering doping concentrations (albeit at the cost of higher driving voltages) or by engineering the guided mode to minimize spatial overlap with the heater. These strategies suggest a viable route to extending the platform across the full long-wave infrared band.

Third, switching speed is ultimately limited by the heater thermal time constant, and can be engineered to reach the microsecond regime for both amorphization and crystallization, faster than LCoS and competitive with or superior to DMDs. However, the passive-matrix addressing used here restricts frame rates, as programming proceeds row-by-row or column-by-column. Realizing the full potential of fast PCM pixels will require pixel-level transistor integration for active-matrix addressing, an advance that our foundry-compatible platform is uniquely positioned to achieve.

In sum, the platform demonstrated here lays the groundwork for a new class of scalable, pixel-addressable active metasurfaces across the mid-IR and other spectral regimes. By combining BEOL-integrated PCMs, foundry-fabricated microheaters, and integrated selectors or transistors within a crossbar architecture, we establish a manufacturing pathway that is directly compatible with high-volume silicon photonics production. This convergence of scalability, broadband operation, and nonvolatile programmability opens opportunities well beyond amplitude-only modulation. With suitable metasurface designs, the same architecture can be adapted for phase, polarization, or orbital angular momentum control, enabling complex wavefront engineering across large apertures. In the longer term, the ability to approach single–meta-atom addressability offers a route toward reconfigurable "universal optics"—devices capable of implementing arbitrary linear optical transformations on demand. Coupled with advances in endurance, pixel density, and active-matrix driving, such metasurfaces could form the basis of compact, high-speed, and multifunctional systems for applications spanning beam steering, adaptive imaging, free-space communications, and dynamic scene projection.

**Table 1. Summary of representative literature reports on electrically addressed reconfigurable metasurfaces.** *N/A* indicates that the corresponding information was not reported in the referenced work.

| Reference | Mechanism | Array architecture | Operation wavelength | Operation mode | Nonvolatile | Pixel switching time/speed | Endurance |
|---|---|---|---|---|---|---|---|
| *Nat. Commun.* **10**, 3654 (2019)[10] | Quantum-confined Stark effect in multi-quantum wells | 1D fanout | ~ 920 nm | Reflective | N | < 1 μs | N/A |
| *Science* **364**, 1087-1090 (2019)[11] | Liquid crystal | 1D fanout | ~ 650 nm | Transmissive | N | N/A | N/A |
| *Nat. Nanotechnol.* **16**, 69-76 (2020)[12] | Carrier injection in indium tin oxide (ITO) | 1D fanout | 1.34 μm and 1.56 μm | Reflective | N | 5.4 MHz (3-dB bandwidth) | N/A |
| *Nat. Commun.* **13**, 5811 (2022)[13] | Electromechanical actuation | 1D fanout | ~ 1.52 μm | Reflective | N | N/A | N/A |
| *Light Sci. Appl.* **11**, 141 (2022)[14] | Liquid crystal | 1D fanout | 460 nm – 650 nm | Reflective | N | N/A | N/A |
| *ACS Nano* **17**, 16952-16959 (2023)[15] | Liquid crystal | 1D fanout | ~ 650 nm | Reflective | N | 0.8 ms (switch on) 1.2 ms (switch off) | N/A |
| *Nano Lett.* **24**, 9961-9966 (2024)[17] | Electrochemical modulation | 1D fanout | 633 nm | Reflective | N | 40 ms (switch on) 55 ms (switch off) | N/A |
| *Nat. Commun.* **11**, 3574 (2020)[19] | Liquid crystal | 2D fanout | 633 nm | Reflective | N | 65 ms (switch on) 40 ms (switch off) | N/A |
| *Nanophotonics* **11**, 2719-2726 (2022)[18] | Carrier injection in ITO | 2D fanout | ~ 1.3 μm | Reflective | N | N/A | N/A |
| *Nat. Nanotechnol.* **16**, 661-666 (2021)[37] | PCM | Single pixel | ~ 1.55 μm | Reflective | Y | 5 μs (amorphization) 500 ms (crystallization) | ≥ 40 |
| *Nat. Nanotechnol.* **16**, 667-672 (2021)[36] | PCM | Single pixel | ~ 700 nm | Reflective | Y | 520 ns (amorphization) 21 μs (crystallization) | ≥ 100 |
| *Nat. Commun.* **13**, 1696 (2022)[35] | PCM | Single pixel | ~ 1.5 μm | Reflective | Y | 200 ns (amorphization) 200 μs (crystallization) | ≥ 50 |
| *Nano Lett.* **25**, 7435-7441 (2025)[38] | PCM | Single pixel | ~ 1.2 μm | Reflective | Y | 8 μs (amorphization) 16 ms (crystallization) | 6 |
| *Adv. Mater.* **36**, 2400627 (2024)[34] | PCM | Single pixel | ~ 1.5 μm | Transmissive | Y | 10 μs (amorphization) 2 s (crystallization) | 1,250 |
| *ACS Nano* **18**, 11245-11256 (2024)[16] | PCM | 1D fanout | ~ 1.52 μm | Transmissive | Y | 1 μs (amorphization) 80 μs (crystallization) | ≥ 1,000 |
| This work | PCM | 2D crossbar with pixel-integrated selectors | ~ 2.6 μm | Transmissive | Y | 13 μs (amorphization) 11 ms (crystallization) | 16,700 |

## Methods

**Device fabrication and packaging.** The BEOL device fabrication process is depicted in Fig. S1. Foundry-fabricated dies were first mounted on 4-inch Si carrier wafers to facilitate handling during processing. Each die consisted of a 220-nm moderately doped Si layer with heavily doped $n^{++}$ contact wings on a 2-µm buried oxide (BOX). To define the heaters, ~ 5 µm of AZ nLOF 2035 resist was spin-coated, prebaked at 113 °C for 60 s, exposed in an MLA Heidelberg 150 at 375 nm with a dose of 500 mJ/cm$^2$, post-exposure baked at 113 °C for 60 s, and developed for ~85 s in AZ 300 MIF. The patterned resist was ultraviolet (UV) flood-exposed and hard-baked at 125 °C for 10 min to improve resilience during subsequent etching. The chips were then mounted with Santovac oil onto 8-inch Si carrier wafers for thermal conduction and etched in two 5-min steps in an AMAT Precision 5000 system. After this etch, the resist was stripped and a second lithography step defined regions to expose the underlying nitride. The chips were immersed in buffered oxide etch (BOE) until a uniform coloration indicated that the nitride was reached. A third lithography step was then performed to expose the moderately doped Si heater region while protecting the $n^{++}$ wings. In this step, 220 nm of $SiN_x$ and 100 nm of $SiO_2$ were etched away.

Following heater definition, a negative resist mask was patterned for rectangular GSST pads extending over the heater region. GSST thin films were prepared using thermal evaporation from a single $Ge_2Sb_2Se_4Te$ source. Bulk starting material was synthesized using a standard melt quench technique from high-purity (99.999%) raw elements[41]. 220 nm of GSST was deposited by thermal evaporation using a custom-designed system (PVD Products, Inc.) at a base pressure of $2 \times 10^{-6}$ Torr with a deposition rate of ~ 1 nm/s. Details of the deposition protocols are discussed elsewhere[42]. Lift-off was performed overnight in N-methyl-2-pyrrolidone (NMP). For metasurface patterning, 500 nm of ZEP 520A was spin-coated and baked at 180 °C for 2 min, followed by a conductive ESpacer 300 layer. Windows were written using electron beam lithography, compensating for the ~ 5 µm topography between the heater surface and surrounding regions. After development in ZED 50 N for 90 s, the GSST in the exposed windows was etched in a SAMCO 230iP ICP-RIE with 25 standard cubic centimeters per minute (sccm) $SF_6$ and 15 sccm Ar at 0.5 Pa, using 100 W bias and 50 W inductively coupled plasma (ICP) power. Etching was performed in short steps with intermittent inspection to ensure complete removal of the GSST without over-etching into the Si heaters. The hardened ZEP mask was stripped by overnight soaking in NMP, followed by gentle sonication.

Encapsulation of the GSST metasurface was performed in two stages. First, 20 nm of $Al_2O_3$ was deposited by atomic layer deposition (ALD) at 150 °C in a Veeco Savannah 200. A thick encapsulation layer was then added by reactive sputtering in an AJA ATC Orion 5 system, with a base pressure of $2.2 \times 10^{-6}$ Torr. Two Si targets were powered at 100 W each, with 7 sccm Ar and 5 sccm $N_2$ at 3 mTorr, yielding ~ 700 nm of $SiN_x$. Windows were opened back to the contact pads using photolithography and a standard $SiN_x$ etch recipe in the SAMCO 230iP. To suppress background transmission through oxide regions outside the heaters, apertures were deposited by electron beam evaporation of 20 nm Ti / 120 nm Pd using the same AJA tool. Pd was chosen for its high melting point and low coefficient of thermal expansion, ensuring robustness under thermal cycling.

Finally, chips were wire-bonded to custom PCBs with an MEI 1204D bonder. Packaged devices were secured in custom 3D-printed holders designed to allow direct backside optical illumination during characterization.

**Optical simulation.** The metasurface optical response was modeled using RCWA implemented with the open-source MATLAB® package RETICOLO V8[43]. Simulations were performed at normal incidence with the incident polarization aligned along the *y*-direction, and illumination applied from the backside through the Si substrate. The simulated stack comprised the Si substrate, a 2-μm buried BOX layer, a 220-nm Si device layer, and a 220-nm GSST fishnet metasurface. The GSST was encapsulated by 20 nm of $Al_2O_3$ and further overcoated with 700 nm of $SiN_x$, with air as the top cladding.

**Device characterization.** The devices were characterized using the setup shown in Fig. 4a. The chip was illuminated using a tunable Firefly M2 laser source. For spectral measurements, the wavelength was swept in 1-nm steps with a 0.1 s dwell time. The sweep was manually synchronized with a MATLAB routine that simultaneously collected images from a Telops IR SPARK M150 camera operating in radiometric temperature mode, scaled between 18 °C and 100 °C. The sample image was focused onto the camera, and the relative positions of the chip and camera were adjusted to achieve the desired magnification.

## Data availability

The data that support the findings of this study are available from the corresponding authors upon reasonable request.

## Acknowledgments

This work was funded by the Air Force SBIR Program under contract FA2394-23-C-5076, the National Science Foundation under awards 2329088 and 2132929, and National Research Foundation of Korea (No. RS-2025-00515651 and No. RS-2025-02213804). B.M. acknowledges support provided by the Draper Scholar Program. This work was carried out in part through the use of MIT.nano's facilities. The authors cordially acknowledge Dr. Anu Agarwal and Prof. Lionel C. Kimerling for access to characterization instruments.

## Author contributions

C.C.P. performed device design, fabrication and characterization. M.P. assisted in device fabrication. O.M. and H.B. contributed to material and deposition process development. R.S. and K.R. prepared the bulk source materials. R.C. helped with device design. K.P.D. performed thermal modeling. L.R. and B.M. contributed to etching process development. H.B., K.R., A.M., H.J.K., D.C., T.G., and J.H. coordinated and supervised the research. C.C.P. and J.H. drafted the manuscript. All authors contributed to technical discussions and preparing the paper.

## Competing financial interests

The authors declare no competing financial interests.